# GPTFF：A high-accuracy out-of-the-box universal AI force field for arbitrary inorganic materials

Fankai Xie, Tenglong Lu, Sheng Meng*, Miao Liu*

Beijing National Laboratory for Condensed Matter Physics, Institute of Physics, Chinese Academy of Sciences, Beijing 100190, China

School of Physical Sciences, University of Chinese Academy of Sciences, Beijing 100190, China

Songshan Lake Materials Laboratory, Dongguan, Guangdong 523808, China

Center of Materials Science and Optoelectronics Engineering, University of Chinese Academy of Sciences, Beijing 100049, China

*Corresponding author: smeng@iphy.ac.cn, mliu@iphy.ac.cn

## Abstract

This study introduces a novel AI force field, namely graph-based pre-trained transformer force field (GPTFF), which can simulate arbitrary inorganic systems with good precision and generalizability. Harnessing a large trove of the data and the attention mechanism of transformer algorithms, the model can accurately predict energy, atomic forces, and stress with Mean Absolute Error (MAE) values of 32 meV/atom, 71 meV/Å, and 0.365 GPa, respectively. The dataset used to train the model includes 37.8 million single-point energies, 11.7 billion force pairs, and 340.2 million stresses. We also demonstrated that GPTFF can be universally used to simulate various physical systems, such as crystal structure optimization, phase transition simulations, and mass transport.

# Introduction

Molecular dynamics (MD) solves Newton's equations of motion to monitor particle positions and velocities as a function of time, and can simulate phenomena such as protein folding, chemical reactions, phase transitions, etc., offering insights into the behaviors of materials from the atomistic level[1-6]. The method is limited primarily to the accuracy of the force fields, which represent how the atoms interact with each other at the microscopic scale. Polishing the force fields is a central topic for the MD community.

The MD method is presently undergoing a drastic transformation when Artificial Intelligence (AI) and high-throughput computation are introduced: the AI allows the force field to have several magnitudes more parameters to outperform the traditional analytic force field function form; the high-throughput computation creates the dataset for AI model training process, level-up the accuracy of the force fields[7, 8]. Recently, the raising of AI force fields has improved the accuracy and applicability of force fields notably, propelling MD to a stage where simulations can be accelerated by up to $10^6$ times compared to the efficiency of density functional theory (DFT) [9]without sacrificing accuracy too much[10-13]. Hence, a viable roadmap for MD method is acknowledged, and the key challenge lies in effectively training the force field with the larger the better trove of datasets with superior quality.

In the past, significant progress has been achieved in developing the AI force field, which can be divided into three stages. At the early stage, people focused on creating effective mathematical algorithms and descriptors to capture the key physics of interatomic interactions by working with a small dataset. For instance, the SOAP descriptor-based Gaussian Approximation Potential(GAP) [10, 14] is of this kind, and there are many successors such as DeepMD[11], SNAP[15], and NNP[12, 13], etc[16-19]. Subsequently, the emergence of neural networks and the expansion of available datasets propelled the AI force field to an elevated level. This advancement involved an increase in the number of parameters and data, leading to substantial improvements in model accuracy without requiring explicit descriptors. Examples of this stage include MegNet[20], DimeNet[21], SchNet[22] and CGCNN[23], ALIGNN[24]. More recently, models for atomistic materials science have progressed to a universal force field stage, necessitating the incorporation of a significant trove of data from both equilibrium and nonequilibrium states and an efficient training process. The obtained force field at this stage, such as M3GNet[25], CHGNet[26], and PFP[27], ALIGNN-FF[28], can be universally applied to nearly any close-to-equilibrium system.

As indicated by several existing literature, the progress of the AI force field is predominantly driven by advancements in data. Previously, Liang et al. [30] showcased that model accuracy increases marginally with the growth of data points, following a power law, meaning that increasing the size of the dataset by one magnitude can improve the model accuracy twice (reduce to half of the mean average error). Deepmind's GNoME project[29] also supports this observation, suggesting that model improvement can be achievable through the inclusion of more data.

On the other hand, the quality of data is crucial, but it is usually challenging to measure and benchmark the quality of the dataset quantitatively. The Materials Project[31] serves as testimony to this, despite not being the largest computational DFT dataset in the field, it is widely utilized as it has high data quality amongst many available[20, 23, 24] and even larger open source datasets out there[32].

Meanwhile, the AI algorithm itself is rapidly evolving in tandem with the recent advances in vision and language models[33, 34]. Nowadays the algorithm developed for fundamental models now can easily incorporate billions, if not trillions, of parameters[35], therefore implies that the AI force field, which usually has less than 1 million parameters, can be further improved if a sufficiently large, good-quality dataset can be obtained.

In this study, we would like to announce a new AI force field, namely graph-based pretrained transformer force field (GPTFF), aiming to showcase the effectiveness of incorporating a large dataset, encompassing 37.8 million energies, 11.7 billion force pairs, and 340.2 million stresses, and beneficial of leveraging the attention mechanism inherent in transformer algorithms[36]. Through this approach, we have successfully developed a pretrained transformer model capable of accurately predicting the energy, atomic forces, and stress of any given materials system, achieving the Mean Absolute Error (MAE) values of 32 meV/atom, 71 meV/Å, and 0.365 GPa for energies, forces, and stress, respectively. Compared to the existing models, GPTFF can achieve better generalization. Furthermore, we have demonstrated its capacity for solving various physical science problems, such as optimizing structures of arbitrary compounds, simulating phase transition of a metal system under external strain, and investigating mass transportation phenomena in ionic compounds.

## Methods

**Model architecture**

For the model construction, we employ Graph Neural Networks (GNNs)[37-39] to represent crystal structures, as illustrated in Figure 1. Like many pioneer force fields that require the construction of atomic coordinate descriptors to ensure rotational, translational, and permutation symmetries[10, 11, 15], using GNNs to represent crystal structures can easily satisfy these symmetry relationships. Each atom is projected into a high-dimensional space according to its element type and is represented in the form of an embedding vector. The geometric structure of each atom's local environment, such as the bond lengths between atoms, is represented by the edges in the GNN, forming an edge vector. The edge vector is represented by concatenating the node vectors representing the two atoms and the vector of information about the distance between them.

Additionally, we incorporate bond angle information related to three-body interactions into the model, achieved by concatenating the node and edge vectors of the atoms forming the bond angle. This allows the model to further learn the interaction relationships between different atomic nodes and edge vectors, thereby improving the model's predictive power significantly. During the high-dimensional representation process of the bond angle, we directly consider the neural network as a scalar function, thereby mapping the Cosine value of the bond angle from one dimension to high dimensions.

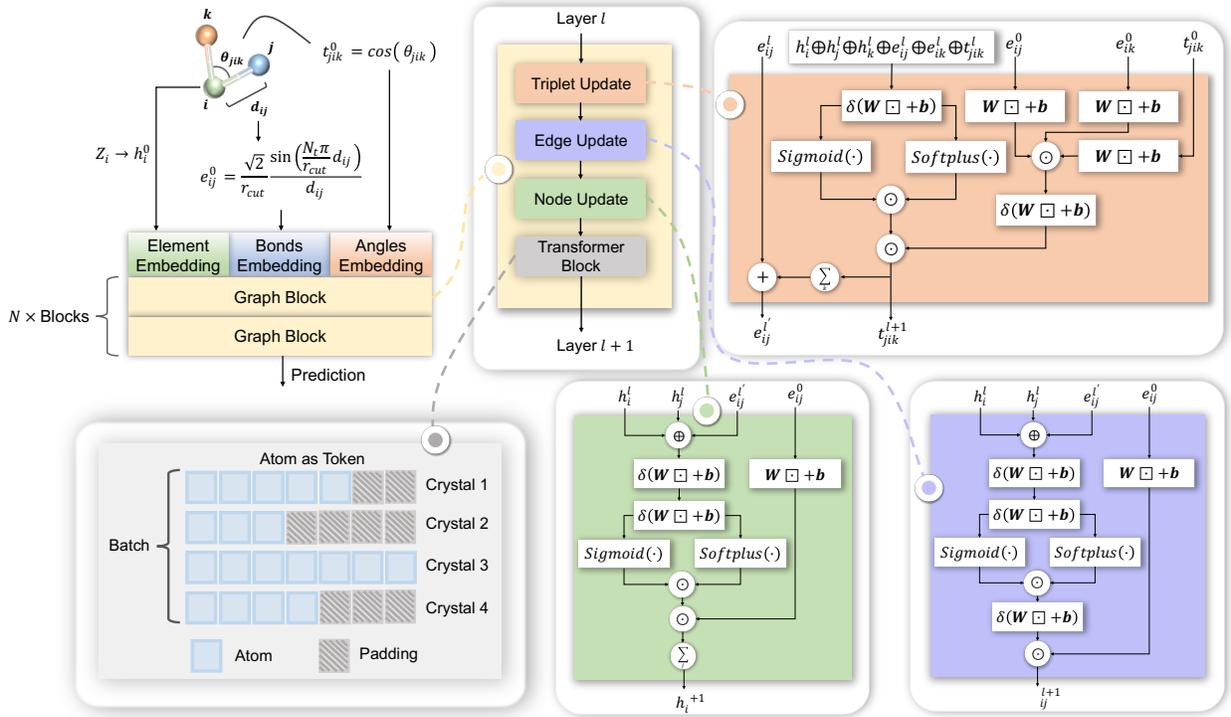

**Figure 1**. The schematic of the model architecture. Information such as element, bond, angles and transformer are represented and updated in graph block.

In the model prediction and inference process, the vector representation within the atomic cutoff radius is predicted by the GNN into the form of atomic energy, and the sum of the atomic energies finally yields the total energy of the system. For the calculation of atomic forces, we use the method of automatic differentiation[40]. The force on an atom can be obtained from the negative gradient of the total energy $E$ with respect to the atomic coordinates, i.e., $\vec{F}_i = \frac{\partial E}{\partial \vec{x}}$. The expression for stress calculation is: $\vec{\sigma} = \frac{1}{V}\frac{\partial E}{\partial \vec{S}}$, where V represents the volume of the system, and $\vec{S}$ represents the strain of the system.

**Dataset**

The training dataset is provided by Atomly.net team, which includes the trajectory of structural optimization of 2,234,767 crystals structures using DFT. All the calculations are conducted utilizing the consistent calculation parameter to yield dataset with good quality, e.g. GGA-PBE[41] is selected as the pseudopotential, the energy cutoff is 520eV, and the 5.4 version of the POTCARs as implemented in the VASP code[42-45] are employed for all the calculations.

Overall, the dataset contains 37.8 million single point energies, 11.7 billion atomistic force vectors, and 340.2 million stresses. Figure 2a displays the statistics of the chance of appearance of each element, and their chance of coappearance of two elements. Within the dataset, there are 349,043 single point energies from the equilibrium state, and 37.4 million single point energies from the nonequilibrium states, roughly 27.6 times larger than the MPtrj dataset[26]. The larger circle size in figure 2 denotes the higher frequency of appearance of the given element. The shade of the strings represents the likelihood they are coexisting in a single compound. Figure 2a demonstrates that the dataset has an evenly distributed sampling of the entire phase space, covering a large and less-biased structural space and chemical space.

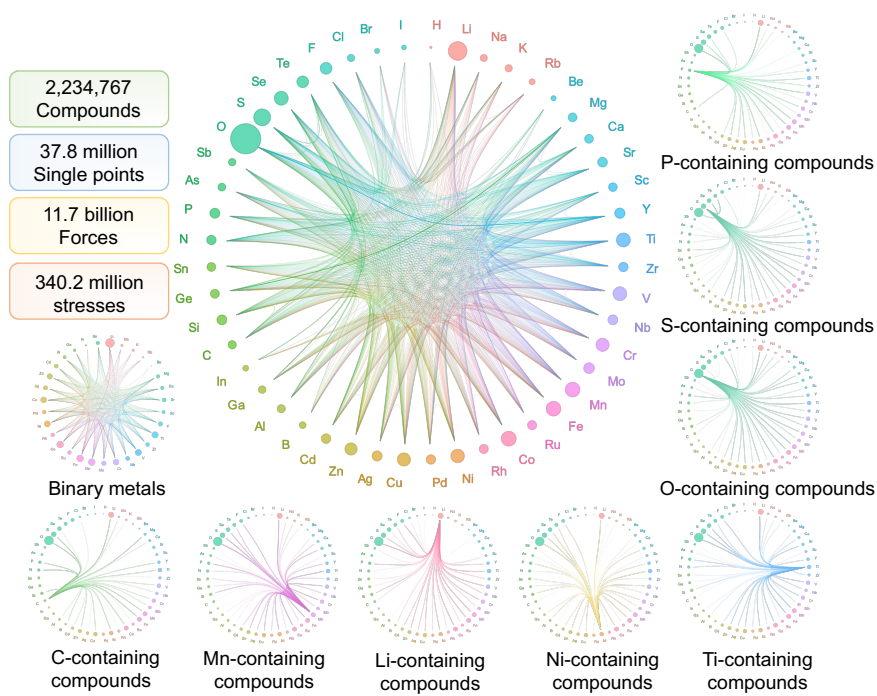

**Figure 2.** Element and training data distribution in atomly database. The training data contains 2,234,767 compounds which generated about 37.8 million single points during structure optimization via DFT calculation.

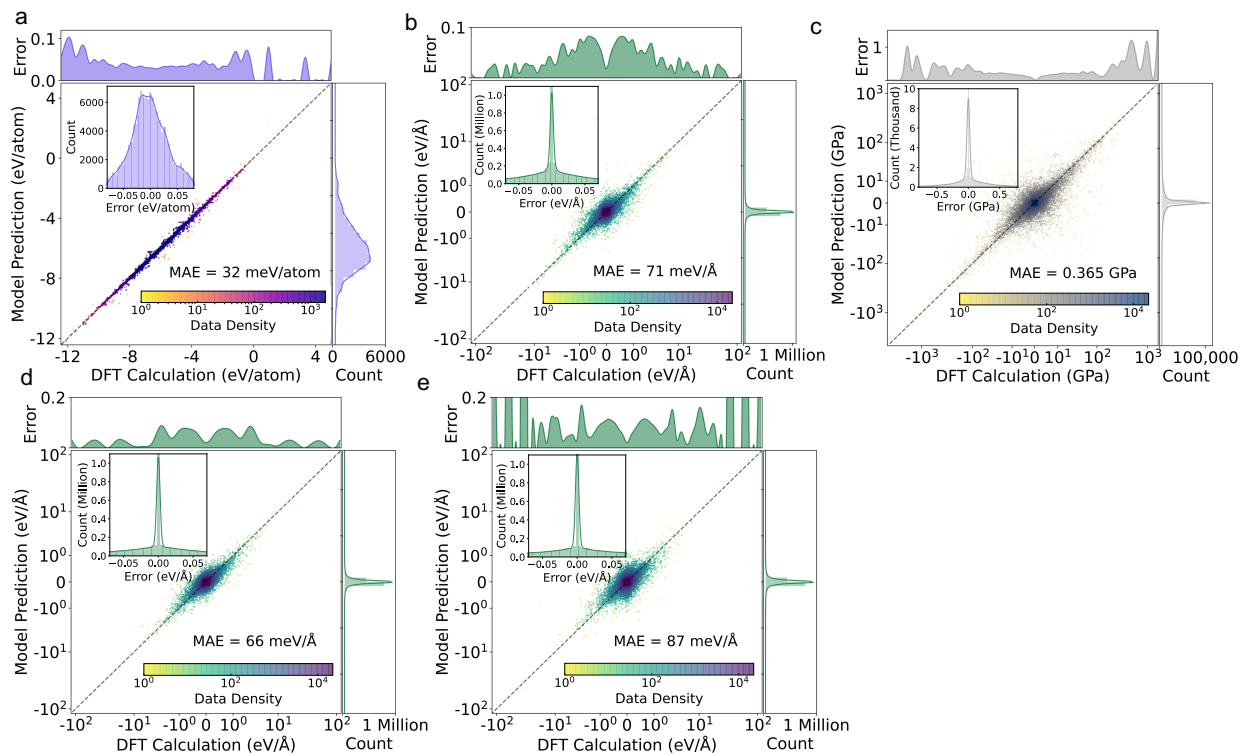

**Figure 2.** a-c. The evaluation MAE results and data distribution of universal force field model on energy, force, stress regression targets in test dataset. d-e. The model performance comparison of our model and CHGNet on structures which not appear in Materials Project. d is the performance of our model, e is the result of CHGNet.

**Training process**

The dataset was divided into training, validation, and test sets. In detail, as dataset is gigantic, we randomly pick up 100,000 data points for validation and test sets separately, for faster validation. It is ensured that the crystal structures in the test set never co-appear in the training and validation sets to yield a valid model. We do not truncate or remove any datapoints in our dataset to ensure that the model have a good generalization.

The AdamW[46] optimization is employed in our model, and a sampling learning method is adopted, where each epoch randomly samples about 1 million data from the training set for model training, with a total of 500 epochs.

Given the inherent limitations of GNNs, there is a saturation point at which increases in model depth and parameters fail to further enhance model performance[47]. To address this issue, we have incorporated Transformer[36] modules into our model. This integration serves to boost the number of parameters, enabling the model to learn a greater amount of latent information within the data. The model currently has 502,465 parameters and can be easily scaled to a larger model with more parameters, such as more than 1 million parameters. Transformers demand a relatively small learning rate during the training process to ensure model stability[48]. Consequently, we adopted an initial learning rate of $2 \times 10^{-4}$ in our training process, which gradually decayed to $5 \times 10^{-6}$ toward the end of the training. This strategy allows for a more controlled and stable training process, minimizing the risk of overshooting the optimal solution and ensuring a robust and reliable model performance.

## Results

**Model performance**

The GPTFF can accurately predict the energies, forces, and stress for any atomistic configuration, hence can serves a valid universal force field. In Figure 3b, it is evident that GPTFF demonstrates an energy error as low as 32 meV/atom, force error of 71meV/Å, and system stress error of 0.365GPa on the test dataset. These results outperform both the M3GNet (MAE = 35 meV/atom) and CHGNet (MAE = 33 meV/atom, without magmom) because our AI force field is built upon a significantly large dataset, substantially enhances the model's generalizability.

For a testing purpose, we conducted additional performance tests on our model and CHGNet for comparison, utilizing a small testing dataset of 16,653 structures, in which all the compounds are new and do not exist in neither the Materials Project nor our training dataset. As depicted in Figure 2c, in this test, the accuracy of CHGNet drops to 87 meV/Å for atomic forces, while our model achieved a higher prediction accuracy of 66 meV/Å.

**Application on structural optimization**

One of the fundamental applications of a universal force field is its ability to quickly optimize any given crystal structure, making it suitable for rapid screening and relaxing unknown structures. It is common practice to create new structures by element substitution, and element substitution will introduce strain in the system as the size of different atom species are not the same, therefore the newly generated structures are not in an equilibrium state, and a structure optimization process is required at this point. GPTFF can be employed to quickly optimize these structures. To demonstrate it, we conducted the calculate the equation of state on the 39 structures using both the DFT and GPTFF. Those 39 compounds cover both ionic compounds and alloys. As shown in Figure 3, the GPTFF can accurately calculate the equation of state of those systems and find the correct equilibrium volume ($R^2$=0.996). It implies that the GPTFF has good precision, and the model can be universally applied to various systems for predict the equilibrium structures and energies.

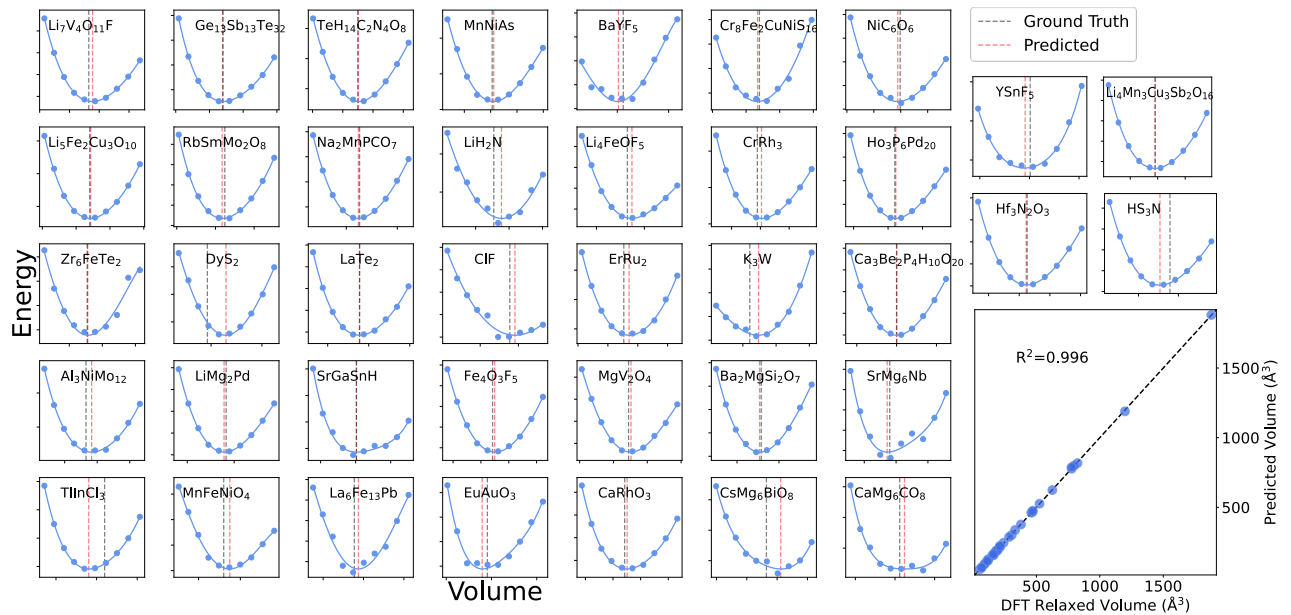

Figure 3. The equation of state results predicted by our universal force field. The model optimizes the structure from initial structure which is not relaxed. The black dashed line represents the final optimized result from DFT, while the red dashed line represents the volume at the equilibrium position after optimization by the force field model.

**Application on MD simulation**

We also conducted tests using the GPTFF to run MD simulations for both metal and ionic compounds. Previously, it was not feasible for empirical MD to easily and accurately simulate ionic compounds as there is no accurate out-of-the-box force field this type compounds[26]. For the metal system, we examined the phase transitions in Titanium (Ti) from HCP to FCC due to stretching along $[10\bar{1}0]$ direction; while for ionic compounds, we calculated the ionic conductivity of the Li cations in $Li_3YCl_6$ system, which is a useful solid-state electrolyte material for lithium ionic batteries.

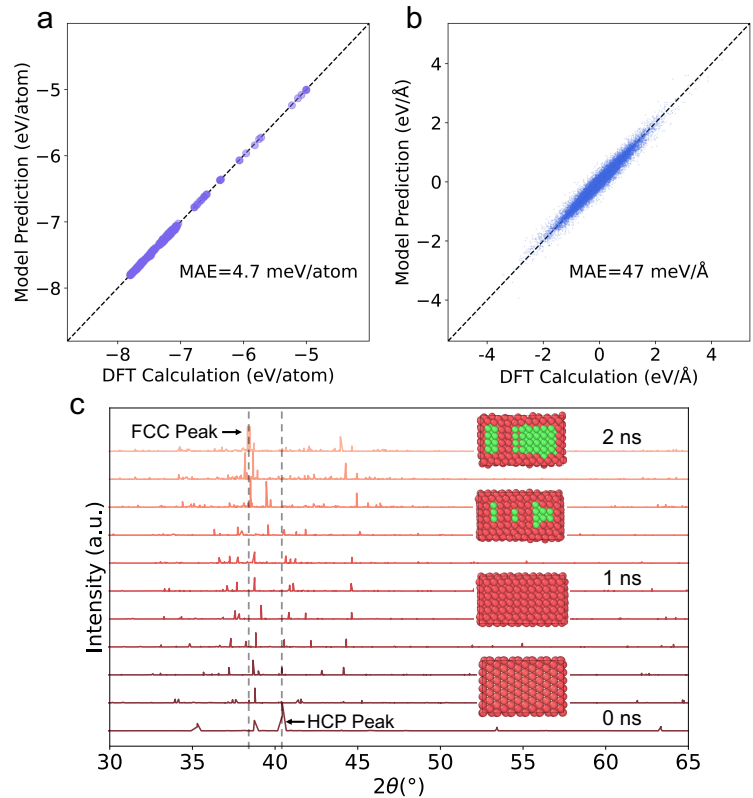

**Figure 4**. a-b. Finetuning results of energy and forces. c. XRD results during MD simulation. With increasing of strain in $[10\bar{1}0]$ direction, HCP→FCC phase transition occurs in Titanium system.

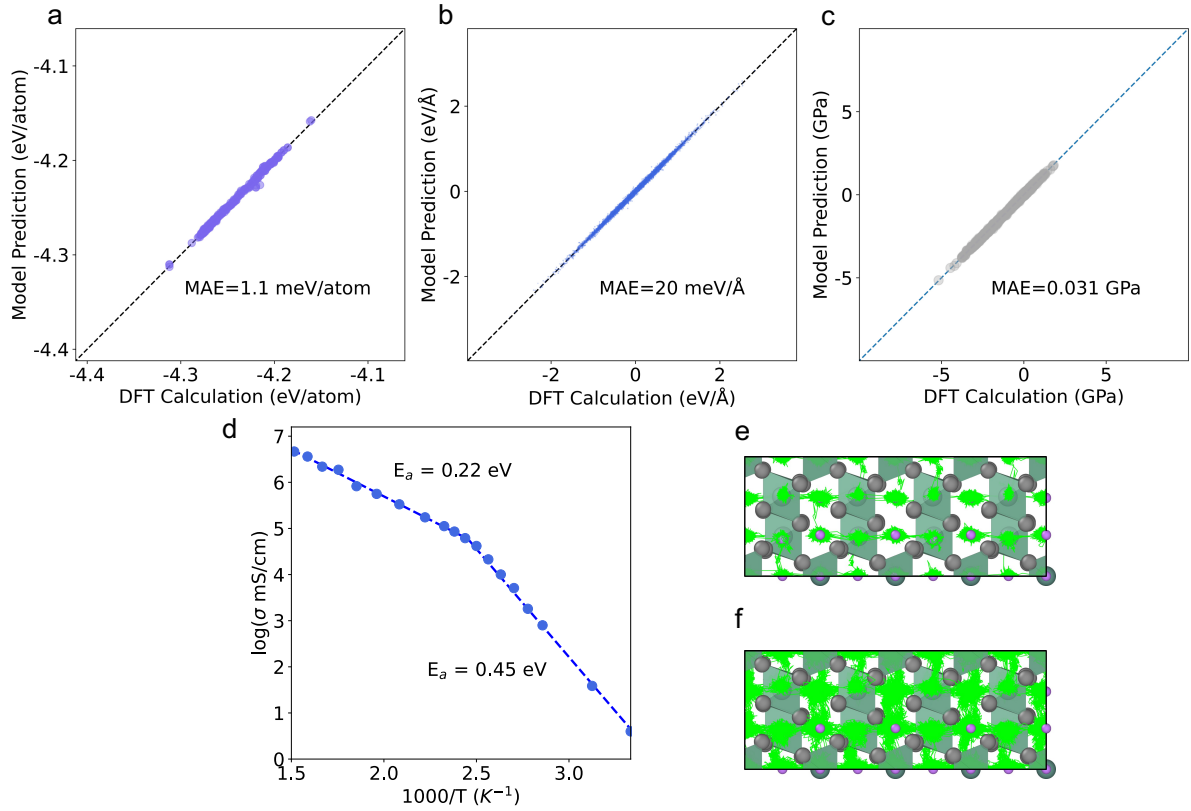

Figure 5. a-c. Finetuning results of energy, force and stress. d. The relationship between temperature and ionic conductivity of Li$_3$YCl$_6$. e-f. Trajectory of lithium ions at 300K and 500K.

There are existing experimental literatures indicating that for the HCP-stacked titanium system, if a tensile strain is applied in the $[10\bar{1}0]$ crystal direction, the $(10\bar{1}0)$ crystal plane slips, thereby generating a $\frac{1}{6}<1\bar{2}10>$ Shockley partial dislocation[49, 50]. The generation of this dislocation exactly transforms the HCP-stacked titanium system into an FCC-stacked titanium system. Accurately capture such HCP-to-FCC phase transition requires a high-precision force field. Previously, the empirical embedded atom potential usually has an issue to distinguish the energies of HCP and FCC as both are close pack structures and the first-shell coordination numbers are the same for both HCP and FCC. Generally, the empirical embedded atom potential has difficulties on distinguish two phases, and therefore a various modified versions of embedded atom potentials were developed to mitigate this issue[51, 52]. However, none of them can reach the accuracy close to that of DFT and capture the HCP-to-FCC phase transition energies correctly.

As shown in Figure 4, by utilizing the GPTFF, HCP-to-FCC phase transition can be very well simulated in MD runs. In the MD simulation, the system has 1,008 atoms. The NVT ensemble is selected at low temperature of ~100K, to avoid thermodynamic disturbances as higher temperatures would vibrate the atoms to blur the peaks on the XRD pattern. Figure 4c shows the snapshot of the crystal structures and XRD peaks during the entire HCP-to-FCC phase transition simulation process. As the MD progresses, an FCC phase peak starts to appear at about 38° on the XRD spectrum, indicating that the occurring of the phase change, which can also be seeing on the atomic structures, in which the red and green regions represent the HCP and FCC stacking respectively.

Fast lithium cation transportation is curial for solid lithium electrolytes in lithium-ion batteries[53]. A typical characteristic of superionic conductors is that within a large temperature range, the diffusion coefficient may follow different Arrhenius equation at different temperature region as the high temperature opens up new diffusion channels, dividing the Arrhenius plots into two zones. To verify the accuracy of the universal force field in ionic systems, we used the force field model to perform molecular dynamics simulations on the $Li_3YCl_6$ system and calculated the ionic conductivity at different temperatures from 300K to 700K. Considering the volume expansion effect as well as the possible phase transitioning brought by temperature, we used the NPT ensemble in the simulation. Since the van der Waals interactions is important, especially in the $Li_3YCl_6$ system[54], we used the optB88[55] corrected DFT to accurately calculate the structural data of a small amount of $Li_3YCl_6$ at different temperatures to obtain a dataset of about 2000 snapshots, and then finetuned the GPTFF. By adding a small amount of data, the pre-trained model can be further finetuned to a much more accurate form a specific system, which is $Li_3YCl_6$ in this case.

Then, we performed molecular dynamics simulations utilizing the finetuned model. Figure 5d shows the ionic conductivity at different temperatures. The ionic conductivity at 300K calculated using the GPTFF, which is 0.6 mS/cm, is in good agreement with the experimental observation, which is 0.51mS/cm[56]. Figure 5 e-f shows that the lithium diffusion channel below transition temperature is mainly concentrated in one dimension. With temperature goes up above the transition point, new diffusion channel of lithium cations opens up in all three dimension. The force field model accurately predicts a superionic transition temperature of 425K. Under temperature range of 230-360K, the activation energy as predicted from the model is 0.45eV, in line with the experimental observations, which is 0.40eV[56]. These results reflect the high accuracy of the GPTFF for calculation the ionic compounds.

## Discussions

The emerge of the universal AI force field has opened a new avenue for theoretical materials science and chemistry. These models facilitate rapid, high-precision theoretical simulations and property predictions for arbitrary systems, reshaping the methodologies that were built on top of traditional DFT computations and empirical force fields. Harnessing the advance of the AI force field, it becomes feasible to conduct high efficiency, high accuracy of MD calculations.

In this work, we have developed the GPTFF and demonstrated its accuracy and convenience in property prediction and atomic scale simulation. Leverage the dataset that covers large chemical and structural phase of inorganic compounds, the GPTFF model endows superior generalization capabilities. In principle, a critical determinant in the training process is not solely the model architecture but, more importantly, the quality of the data employed, suggesting that the dataset is the fundamental of this type of research, hence credit should be given to people who build up the dataset.

Currently, the supervision labels for GPTFF include energy, atomic forces, and system stress, which indeed utilize only a small amount of information generated by the DFT runs, whereas the information about electron-related properties is mostly ignored at the current level of models. In the forthcoming efforts, the community may aim to enhance the robustness of atomistic AI models by incorporating more data and labels. CHGNET has demonstrated that plugin the magnetic moments into the model can level up the predictive power, and this would be a useful strategy.

In the future, we plan to refine the self-attention mechanism within the transformer architecture to include more geometric information and to increase the model parameters. These would progressively make GPTFF more powerful.

## Conclusions

In conclusion, GPTFF can optimize arbitrary crystal structures rapidly and can be used for large-scale atomic simulation, which enables further exploration of larger materials phase spaces for speeding up the discovery of new materials.

## Model/code availability

The code of the GPTFF will be openly released after peer review.

## Acknowledgments

This research is supported by Chinese Academy of Sciences (Grant No. CAS-WX2023SF-0101, ZDBS-LY-SLH007, and XDB33020000, XDB33030100, and YSBR-047), the National Key R&D Program of China (2021YFA0718700), and the National Natural Science Foundation of China (Grand No. 12025407 and 11934003). The computational resource is provided by the Platform for Data-Driven Computational Materials Discovery of the Songshan Lake Materials laboratory.

## References

1. Hansson, T., C. Oostenbrink, and W. van Gunsteren, *Molecular dynamics simulations.* Current Opinion in Structural Biology, 2002. **12**(2): p. 190-196.
2. Hollingsworth, S.A. and R.O. Dror, *Molecular Dynamics Simulation for All.* Neuron, 2018. **99**(6): p. 1129-1143.
3. Hospital, A., et al., *Molecular dynamics simulations: advances and applications.* Advances and Applications in Bioinformatics and Chemistry, 2015. **8**(null): p. 37-47.
4. Durrant, J.D. and J.A. McCammon, *Molecular dynamics simulations and drug discovery.* BMC Biology, 2011. **9**(1): p. 71.
5. Rapaport, D.C., *The art of molecular dynamics simulation*. 2004: Cambridge university press.
6. Binder, K., et al., *Molecular dynamics simulations.* Journal of Physics: Condensed Matter, 2004. **16**(5): p. S429.
7. Unke, O.T., et al., *Machine Learning Force Fields.* Chemical Reviews, 2021. **121**(16): p. 10142-10186.
8. Poltavsky, I. and A. Tkatchenko, *Machine learning force fields: Recent advances and remaining challenges.* The Journal of Physical Chemistry Letters, 2021. **12**(28): p. 6551-6564.

9. Payne, M.C., et al., *Iterative minimization techniques for ab initio total-energy calculations: molecular dynamics and conjugate gradients.* Reviews of Modern Physics, 1992. **64**(4): p. 1045-1097.
10. Bartók, A.P., et al., *Gaussian Approximation Potentials: The Accuracy of Quantum Mechanics, without the Electrons.* Physical Review Letters, 2010. **104**(13): p. 136403.
11. Zhang, L., et al., *Deep Potential Molecular Dynamics: A Scalable Model with the Accuracy of Quantum Mechanics.* Physical Review Letters, 2018. **120**(14): p. 143001.
12. Behler, J. and M. Parrinello, *Generalized Neural-Network Representation of High-Dimensional Potential-Energy Surfaces.* Physical Review Letters, 2007. **98**(14): p. 146401.
13. Behler, J., *Atom-centered symmetry functions for constructing high-dimensional neural network potentials.* The Journal of Chemical Physics, 2011. **134**(7): p. 074106.
14. Bartók, A.P., R. Kondor, and G. Csányi, *Erratum: On representing chemical environments [Phys. Rev. B 87, 184115 (2013)].* Physical Review B, 2017. **96**(1): p. 019902.
15. Thompson, A.P., et al., *Spectral neighbor analysis method for automated generation of quantum-accurate interatomic potentials.* Journal of Computational Physics, 2015. **285**: p. 316-330.
16. Shapeev, A.V., *Moment Tensor Potentials: A Class of Systematically Improvable Interatomic Potentials.* Multiscale Modeling & Simulation, 2016. **14**(3): p. 1153-1173.
17. Novikov, I.S., et al., *The MLIP package: moment tensor potentials with MPI and active learning.* Machine Learning: Science and Technology, 2021. **2**(2): p. 025002.
18. Hernandez, A., et al., *Fast, accurate, and transferable many-body interatomic potentials by symbolic regression.* npj Computational Materials, 2019. **5**(1): p. 112.
19. Drautz, R., *Atomic cluster expansion for accurate and transferable interatomic potentials.* Physical Review B, 2019. **99**(1): p. 014104.
20. Chen, C., et al., *Graph Networks as a Universal Machine Learning Framework for Molecules and Crystals.* Chemistry of Materials, 2019. **31**(9): p. 3564-3572.
21. Gasteiger, J., J. Groß, and S. Günnemann, *Directional Message Passing for Molecular Graphs.* arXiv e-prints, 2020: p. arXiv:2003.03123.
22. Schütt, K.T., et al., *SchNet – A deep learning architecture for molecules and materials.* The Journal of Chemical Physics, 2018. **148**(24): p. 241722.
23. Xie, T. and J.C. Grossman, *Crystal Graph Convolutional Neural Networks for an Accurate and Interpretable Prediction of Material Properties.* Physical Review Letters, 2018. **120**(14): p. 145301.
24. Choudhary, K. and B. DeCost, *Atomistic Line Graph Neural Network for improved materials property predictions.* npj Computational Materials, 2021. **7**(1): p. 185.
25. Chen, C. and S.P. Ong, *A universal graph deep learning interatomic potential for the periodic table.* Nature Computational Science, 2022. **2**(11): p. 718-728.
26. Deng, B., et al., *CHGNet as a pretrained universal neural network potential for charge-informed atomistic modelling.* Nature Machine Intelligence, 2023. **5**(9): p. 1031-1041.
27. Takamoto, S., et al., *Towards universal neural network potential for material discovery applicable to arbitrary combination of 45 elements.* Nature Communications, 2022. **13**(1): p. 2991.
28. Choudhary, K., et al., *Unified graph neural network force-field for the periodic table: solid state applications.* Digital Discovery, 2023. **2**(2): p. 346-355.

29. Merchant, A., et al., *Scaling deep learning for materials discovery.* Nature, 2023. **624**(7990): p. 80-85.
30. Liang, Y., et al., *A universal model for accurately predicting the formation energy of inorganic compounds.* Science China Materials, 2023. **66**(1): p. 343-351.
31. Jain, A., et al., *Commentary: The Materials Project: A materials genome approach to accelerating materials innovation.* APL Materials, 2013. **1**(1).
32. Kirklin, S., et al., *The Open Quantum Materials Database (OQMD): assessing the accuracy of DFT formation energies.* npj Computational Materials, 2015. **1**(1): p. 15010.
33. Dosovitskiy, A., et al. *An Image is Worth 16x16 Words: Transformers for Image Recognition at Scale*. 2020. arXiv:2010.11929.
34. Devlin, J., et al., *BERT: Pre-training of Deep Bidirectional Transformers for Language Understanding.* arXiv e-prints, 2018: p. arXiv:1810.04805.
35. Touvron, H., et al., *Llama 2: Open Foundation and Fine-Tuned Chat Models.* arXiv e-prints, 2023: p. arXiv:2307.09288.
36. Vaswani, A., et al., *Attention Is All You Need.* arXiv e-prints, 2017: p. arXiv:1706.03762.
37. Zhou, J., et al., *Graph neural networks: A review of methods and applications.* AI Open, 2020. **1**: p. 57-81.
38. Wu, Z., et al., *A Comprehensive Survey on Graph Neural Networks.* IEEE Transactions on Neural Networks and Learning Systems, 2021. **32**(1): p. 4-24.
39. Scarselli, F., et al., *The Graph Neural Network Model.* IEEE Transactions on Neural Networks, 2009. **20**(1): p. 61-80.
40. Paszke, A., et al., *PyTorch: An Imperative Style, High-Performance Deep Learning Library.* arXiv e-prints, 2019: p. arXiv:1912.01703.
41. Perdew, J.P., K. Burke, and M. Ernzerhof, *Generalized Gradient Approximation Made Simple.* Physical Review Letters, 1996. **77**(18): p. 3865-3868.
42. Kresse, G. and J. Hafner, *Ab initio molecular dynamics for liquid metals.* Physical Review B, 1993. **47**(1): p. 558-561.
43. Kresse, G. and J. Hafner, *Norm-conserving and ultrasoft pseudopotentials for first-row and transition elements.* Journal of Physics: Condensed Matter, 1994. **6**(40): p. 8245.
44. Kresse, G. and J. Furthmüller, *Efficiency of ab-initio total energy calculations for metals and semiconductors using a plane-wave basis set.* Computational Materials Science, 1996. **6**(1): p. 15-50.
45. Kresse, G. and J. Furthmüller, *Efficient iterative schemes for ab initio total-energy calculations using a plane-wave basis set.* Physical Review B, 1996. **54**(16): p. 11169-11186.
46. Loshchilov, I. and F. Hutter, *Decoupled Weight Decay Regularization.* arXiv e-prints, 2017: p. arXiv:1711.05101.
47. Ying, C., et al., *Do Transformers Really Perform Bad for Graph Representation?* arXiv e-prints, 2021: p. arXiv:2106.05234.
48. Liu, L., et al., *Understanding the Difficulty of Training Transformers.* arXiv e-prints, 2020: p. arXiv:2004.08249.
49. Yang, J.X., et al., *Proposed mechanism of HCP →FCC phase transition in titanium through first principles calculation and experiments.* Scientific Reports, 2018. **8**(1): p. 1992.
50. Bai, F., et al. *Study on Phase Transformation Orientation Relationship of HCP-FCC during Rolling of High Purity Titanium*. Crystals, 2021. **11**, DOI: 10.3390/cryst11101164.

51. Baskes, M.I. and R.A. Johnson, *Modified embedded atom potentials for HCP metals.* Modelling and Simulation in Materials Science and Engineering, 1994. **2**(1): p. 147.
52. Kim, Y.-M., B.-J. Lee, and M.I. Baskes, *Modified embedded-atom method interatomic potentials for Ti and Zr.* Physical Review B, 2006. **74**(1): p. 014101.
53. Zhang, Z., et al., *New horizons for inorganic solid state ion conductors.* Energy & Environmental Science, 2018. **11**(8): p. 1945-1976.
54. Qi, J., et al., *Bridging the gap between simulated and experimental ionic conductivities in lithium superionic conductors.* Materials Today Physics, 2021. **21**: p. 100463.
55. Klimeš, J., D.R. Bowler, and A. Michaelides, *Chemical accuracy for the van der Waals density functional.* Journal of Physics: Condensed Matter, 2010. **22**(2): p. 022201.
56. Asano, T., et al., *Solid Halide Electrolytes with High Lithium-Ion Conductivity for Application in 4 V Class Bulk-Type All-Solid-State Batteries.* Advanced Materials, 2018. **30**(44): p. 1803075.